\documentclass[graybox]{svmult}

\usepackage{mathptmx}       
\usepackage{helvet}         
\usepackage{courier}        
\usepackage{type1cm}        
\usepackage{makeidx}         
\usepackage{graphicx}        
\usepackage{multicol}        
\usepackage[bottom]{footmisc}

\usepackage{subfig}

\makeindex             

\begin{document}

\title*{Interbank markets and multiplex networks: centrality measures and statistical null models}
\author{Leonardo Bargigli, Giovanni di Iasio, Luigi Infante, Fabrizio Lillo and Federico Pierobon}
\institute{Leonardo Bargigli \at Dipartimento di Scienze per l'Economia e l'Impresa, Universit\'a di Firenze, \email{leonardo.bargigli@unifi.it}
\and Giovanni di Iasio \at  Bank of Italy,  \email{giovanni.diiasio@bancaditalia.it}
\and Luigi Infante \at  Bank of Italy,  \email{luigi.infante@bancaditalia.it}
\and Fabrizio Lillo \at Scuola Normale Superiore, piazza dei Cavalieri 7, Pisa, Italy \email{fabrizio.lillo@sns.it}
\and Federico Pierobon \at European Central Bank, \email{federico.pierobon@ecb.europa.eu}}
%
%
\maketitle

\abstract*{Each chapter should be preceded by an abstract (10--15 lines long) that summarizes the content. The abstract will appear \textit{online} at \url{www.SpringerLink.com} and be available with unrestricted access. This allows unregistered users to read the abstract as a teaser for the complete chapter. As a general rule the abstracts will not appear in the printed version of your book unless it is the style of your particular book or that of the series to which your book belongs.
Please use the 'starred' version of the new Springer \texttt{abstract} command for typesetting the text of the online abstracts (cf. source file of this chapter template \texttt{abstract}) and include them with the source files of your manuscript. Use the plain \texttt{abstract} command if the abstract is also to appear in the printed version of the book.}

\abstract{The interbank market is considered one of the most important channels of contagion. Its network representation, where banks and claims/obligations are represented by nodes and links (respectively), has received a lot of attention in the recent theoretical and empirical literature, for assessing systemic risk and identifying systematically important financial institutions. Different types of links, for example in terms of maturity and collateralization of the claim/obligation, can be established between financial institutions. Therefore a natural representation of the interbank structure which takes into account more features of the market, is a multiplex, where each layer is associated with a type of link. In this paper we review the empirical structure of the multiplex and the theoretical consequences of this representation. We also investigate the betweenness and eigenvector centrality of a bank in the network, comparing its centrality properties across different layers and with Maximum Entropy 
null models.}

\section{Introduction}

In part as a consequence of the crisis burst after the Lehman event in late 2008, financial networks are gaining popularity across policymakers, regulators and academics interested in systemic risk analysis. Networks are useful and natural tools to identify critical financial institutions as well as to understand how distress could propagate within the financial system.\footnote{See for instance \cite{abbassi2013, Allen2000,battiston2007, battiston2012, RePEc:fip:fednsr:354,boss2004network, caccioli2011, cont2011network, fricke2013, gai2010contagion, iazzetta2009topology, iori2008network, iori2006, marsili2012, martinezjar2012, soramaki2007topology}} The type of financial network considered in this paper is the one of direct exposures between financial institutions, which include credit relations, derivatives transactions, cross-ownerships, etc.. 

\begin{figure}[t]
\begin{center}
\includegraphics[width = 0.58\textwidth,keepaspectratio=true]{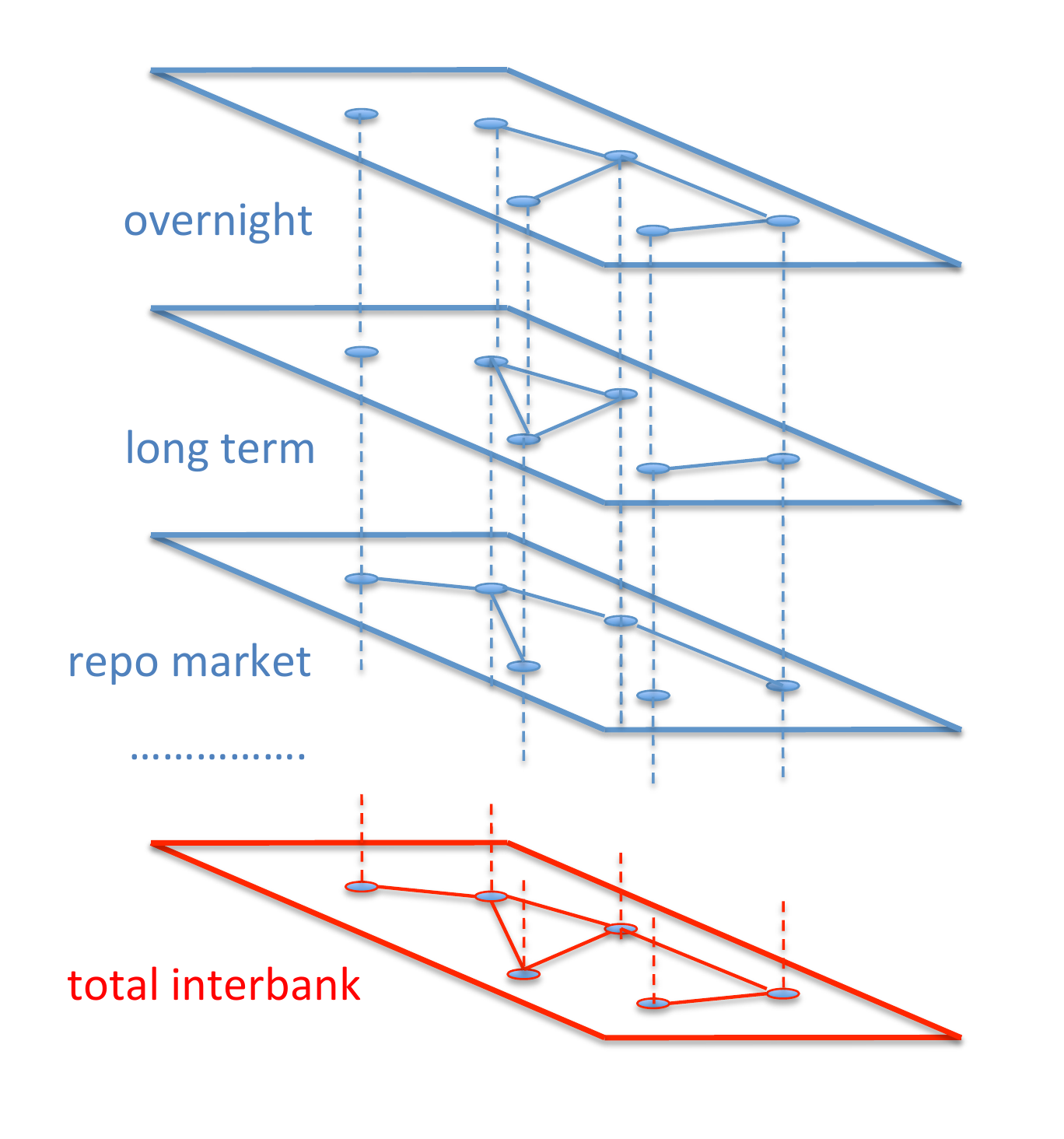}
\end{center}
\caption{Stylized representation of the multiplex structure of the interbank market. Each node is a bank, and links represent credit relations. A layer is the set of all credit relations of the same type. The network in red is the total interbank market, obtained by aggregating all the layers.}
\label{multiplex}
\end{figure}

Interlinkages between any two financial institutions is more complex than the information that can be summarized in a single number (the weight of the link) and a direction, such as in a directed and weighted network. This is due to the fact that between two institutions there exists a multiplicity of linkages, each of them related to one class of claims/obligations. The interplay between different types of relation could be relevant for systemic risk analysis. In the network jargon, such a situation is best modeled with a {\it multiplex network} or simply {\it multiplex}. A multiplex is made up with several "layers", each of them composed by all relations of the same type and modeled with a simple (possibly weighted and directed) network. Since the nodes in each layer are the same, the multiplex can be visualized as a stack of networks or equivalently by a network where several different types of links can coexist between two nodes, each type corresponding to a layer.   Figure \ref{multiplex} shows a 
stylized representation of the interbank multiplex, where three layers are explicitly shown, together with the aggregated interbank network.

The study of financial multiplex  is still at its infancy \cite{montagna2013multi,Bargigli2014}. This paper builds on the analysis performed recently by the authors on the Italian Interbank Market (IIN) (see \cite{Bargigli2014}). Taking advantage of a unique dataset collected by Banca d'Italia, \cite{Bargigli2014} analyze in detail the evolution of the multiplex during the crisis (2008-2012).\footnote{\cite{Bargigli2014}  compare the topological proprieties of each layer, study the similarity between layers, i.e. how much one can learn from a layer knowing another one, and use a Maximum Entropy approach to investigate which high order topological properties of each layer can be explained by the statistical properties of degree and strength. In other words, the paper investigates how much the huge heterogeneity in degree and strength observed in interbank markets significantly constrains (e.g. for the assortativity) or is a sufficient statistics (e.g. for the core-periphery structure) other topological 
properties studied in the recent interbank network literature.} 

In this paper, after a brief description of the [Bargigli et al. (2014)] main findings on the multiplex representation of the interbank market, we add on a novel result by investigating an important network property, previously not considered, namely the network centrality of the banks. Node centrality is critical to identify the most important nodes in the network architecture. There are several different definitions of centrality, depending on the meaning given to the word "important" above. Here we shall consider betweenness centrality and eigenvector centrality, two widespread centrality metrics. Our main research questions can be summarized as follows: (i) Given a centrality measure of a node, how much is it correlated with degree and strength of the node? Is this correlation layer specific? (ii) Is the centrality (absolute or rank) of a bank approximately the same in each layer? Or are there banks which are very central in some layers but essentially peripheral in the others? (iii) How much the 
centrality of a node can be explained by a statistical null model that preserves nodes' degree and strength? Is the centrality ranking  mostly determined by the degree (or strength) ranking?  

The paper is organized as follows. In Section \ref{data} we present the dataset and introduce the construction of our multiplex. In Section \ref{review} we review the main findings of \cite{Bargigli2014} about the Italian Interbank Network. Section \ref{centralitydef} introduces several types of centrality metrics and presents our results, both on the investigation of the centrality property of a bank in each layer and on the comparison with Maximum Entropy null models. Section \ref{conclusions} concludes.

\section{Data description}\label{data}

Our analysis investigates a unique database of interbank transactions based on the supervisory reports transmitted to Banca d'Italia by all institutions operating in Italy (see \cite{Bargigli2014} for details). We focus only on domestic links, i.e. transactions between Italian banks. Information refers to end-of-year outstanding balances at the end of 2008, 2009, 2010, 2011, and 2012. This was a particularly delicate period for the Italian banking system because of the concurrent sovereign debt crisis hardly hitting the Euro zone and Italy in particular. Most banks operate in Italy through a large set of subsidiaries. Distinguishing between intragroup and intergroup transactions is therefore crucial. Since interbank lending and borrowing decisions are normally taken at the parent company level, we assume that the relevant economic agents of the network lie at the group level. We thus focus on data on intergroup transactions consolidated at the group level. As shown in  \cite{Bargigli2014}  this consolidation 
has a dramatic effect on the volumes and the network structure. In fact in the investigated dataset and period the intragroup lending accounts for a fraction of the total volume ranging between 78\% and 89\%. 

We represent the interbank market as a weighted and directed network, where a set of nodes (banks) are linked to each other through different types of financial instruments (edges). The direction of the link goes from the bank $i$ having a claim to the bank $j$ , and the weight is the amount (in millions of Euros) of liabilities of $j$ towards $i$.  

One specificity of our database is the availability of other information on the interbank transactions, namely the maturity and the presence of a collateral. As in \cite{Bargigli2014} we use this information to build a multiplex representation of the interbank network, where each layer describes the network of a specific type of credit relation. Specifically, in terms of maturity we consider
\begin{itemize}
\item Overnight (OVN) transactions
\item Short term (ST) transaction, namely those with maturity up to 12 months excluding overnight
\item Long term (LT) transaction, namely those with maturity of more that 12 months
\end{itemize}
Considering collateralization we distinguish
\begin{itemize}
\item Unsecured (U) loans, i.e. without collateral
\item Secured (S) loans, i.e. with collateral
\end{itemize}
Thus, for example, the symbol $U~ST$ stands for unsecured short term contracts. We have in total 5 possible combinations, since there are no secured overnight transactions. It is important to stress that our data on secured transactions only refer to OTC contracts, while secured transactions taking place on regulated markets and centrally cleared are not included in our database.

\section{The multiplex structure of the Italian interbank market}\label{review}

In  \cite{Bargigli2014} we presented one of the first in-depth analyses of the multiplex structure of an interbank market, investigating in particular the Italian market. Here we summarize our main findings. The first important observation is that interbank market is dominated (in volume) by intragroup lending. To give an idea, in 2012 almost $78\%$ of the volume of the Italian interbank market was traded between two banks belonging to the same group, while only $22\%$ was intergroup lending. Since intergroup lending is the main channel of systemic risk propagation, in  \cite{Bargigli2014}  and here we aggregate banks belonging to the same group and we consider the network of banking groups. 

When we consider the volume in the different layers we observed that in 2012 the layer with the largest activity was $U~LT$ with roughly $41\%$, followed by OVN with $28\%$ and $U~ST$ with $18\%$. The  two layers describing secured transactions have small volumes, namely $9\%$ and $2\%$ in the $S~ST$ and  $S~LT$, respectively. This is mostly due to the fact that today the majority of collateralized trades are operated through Central Counterparties (CCPs) and these trades are not included in our dataset. We found that the three unsecured layers are composed by many nodes, while the two secured one are much smaller. In 2012 the Italian multiplex had 533 banks, and almost all of them were active in the three unsecured layers. In all layers there is a strong correlation between the degree of a node (i.e. the number of counterparts) and the its strength (the total volume lent or borrowed). The Spearman correlation between these two quantities ranges between $0.49$ and $0.94$. Moreover degree and strength are 
also very correlated ($> 0.5$) with bank size (as measured by the total asset). This means that the size of a bank is an important determinant of the amount traded in the interbank market and of the number of counterparts. 

In \cite{Bargigli2014} we compared also the topology of each layer individually. All the layers display a scale free property, i.e. the degree distribution has a power law tail characterized by a tail exponent between $1.8$ and $3.5$. This indicates a very strong heterogeneity of the system, with few important hubs and many low-degree nodes. Each layer has a small diameter, it is strongly reciprocal, and has a large clustering coefficient. Finally, it presents a disassortative mixing and has a clear core periphery structure. Below we will discuss on the relevance of these two related last findings.

The multiplex structure of the system raises immediately the question of the similarity between the layers, not in terms of their generic statistical properties, but on a link-by-link basis. In other words, in \cite{Bargigli2014} we asked how much the presence of a link between two banks in a layer is predictive of the presence of a link between the same two banks in the other layers. Interestingly, the answer is that layers are quite different one from each other, and the knowledge of the links in one layer gives little information about the presence of links in the other layers. On average across pairs of layers, the similarity, as measured by the Jaccard similarity, was roughly $17\%$ in 2012.  This means that by observing a link in a layer, one can predict that the link exists in another layer only in $17\%$ of the cases. This low level of similarity, combined with the high level of similarity between the {\it same} layer in consecutive years (roughly $70\%$), indicates that banks diversify their 
counterparts across different layers, but maintain stable relationships with the same counterparts in a given layer across the years. From a systemic risk point of view, the dissimilarity between layers tells us that the propagation of contagion can be significantly faster when one considers the multiplex structure, as compared with the contagion in individual layers. In fact, simple diffusion models on multiplex networks \cite{Gomez13} show that diffusion can be very fast when the layers are "orthogonal" one to each other. In the second part of this paper we will consider a related question, namely how much centrality measures of a node are specific of a layer or whether different nodes are the most central in different layers. 

Finally, \cite{Bargigli2014} compares the properties of the real multiplex network with those obtained under statistical null models, specifically with Maximum Entropy models (see Section \ref{null} below). The Maximum Entropy principle allows to build explicitly probability distributions of graphs, which are maximally random and satisfy some given constraints. In particular we constrained the ensembles to have, on average, the same degree and/or strength sequence as the real network. This choice corresponds to the (weighted) configuration model. The aim of this analysis is to discriminate which high order properties of the real interbank networks are a mere consequence of the heterogeneity of degrees (or strengths) and which are instead genuinely new properties. We focused on disassortativity, reciprocity, and the presence of triadic motifs and we discussed which of these properties can be reproduced by the null model. To give a specific, yet important, example of the application of Maximum Entropy null 
models to interbank networks,  we discuss briefly the case of the core-periphery structure.

Recently there has been an increasing interest toward the core-periphery structure of the interbank network \cite{RePEc:fip:fedcwp:0912,RePEc:kie:kieliw:1759}. The core is defined as a subset of nodes which are maximally connected with other core members, while the periphery is the complementary subset made of nodes with no reciprocal connections but having only connections with the core \cite{borgatti00}. Recently, \cite{RePEc:fip:fedcwp:0912} defined a tiering model in which core members without links with the periphery are penalized. Despite the fact that the two definitions of core-periphery are different, because of the different objective function, in Appendix 2 of \cite{Bargigli2014} we found that the two models are highly correlated, i.e. the identified cores are essentially the same. The key question now is whether core-periphery is a genuine property of the interbank network or if it can be explained by the strong heterogeneity of degree. To this end in \cite{Bargigli2014}  we generated random 
sample from the Maximum Entropy ensemble where we fix either the average degree or the average weight of each node. By comparing the core-periphery subdivision in the real network or in the random samples, we conclude that they are very similar one to each other\footnote{It is worth noticing that this can also be explained by using the result of \cite{2011arXiv1102.5511L}, showing analytically that the subdivision in core and periphery according to the definition of \cite{borgatti00} is entirely determined by the degree sequence.}. Thus core-periphery subdivision (at least by using the aforementioned definitions) is a consequence of the large heterogeneity of degree. Since, as mentioned above, degree and bank size are strongly correlated, we conclude that in great part core-periphery structure is a consequence of the existence of large and small banks.

\section{Centrality measures}\label{centralitydef}

Centrality is a key concept of network theory, originally developed in social network analysis and rapidly developed to other types of networks. Broadly speaking, the centrality of a node (or of an edge) of a network is a measure of the importance of the node, measuring, for example, how influential is a person in a social network, how critical is an element in an infrastructure network, what is the disease spreading capacity of an individual, etc.. Despite being an important concept, the loose definition given above leads to several different proposed centrality measures, each of them able to capture some specific aspects of the concept of centrality. The review of all the centrality measures is beyond the scope of this paper. In the following we will discuss the three measures of node centrality we are going to use in the analysis of the different layers of the IIN.

{\bf Degree centrality.} Degree of a node is an obvious measure of centrality. A large number of links, in fact, is a symptom of the fact that the node is important for the connection of all the nodes which are linked to it. It is a local measure, i.e. it does not take into account the whole network, but only the local neighborhood of the node. Thus, while for small networks degree is a sensible centrality measure, for large networks it can miss important global characteristics of the importance of the node.

{\bf Betweenness centrality.} One of the most popular global centrality measures is the betweenness centrality (often shortened as betweenness in the following). It quantifies how frequently a node acts as a bridge along the shortest path between two other nodes. More formally, betweenness centrality of a node $v$ is computed in the following way: for each pair of vertices $(i,j)$ one identifies the $N_{ij}$ shortest paths between them and computes the number $N_{ij}(v)$ of them that pass through $v$. The betweenness of node $v$ is
\begin{equation}\label{betw}
C_B(v)=\frac{1}{(n-1)(n-2)}\sum_{i,j\ne v}\frac{N_{ij}(v)}{N_{ij}},
\end{equation}
i.e. the average fraction of shortest paths passing through $v$, where the average is taken across all the pairs of vertices. The number of nodes in the network is $n$ and the normalization factor in equation \ref{betw}, and used in this paper, holds for directed graphs. Intuitively $C_B(v)$ measures how frequently a shortest path between two nodes passes through a given node. 

{\bf Eigenvector centrality.} Eigenvector centrality is defined in terms of the adjacency matrix $A=\lbrace a_{ij} \rbrace$, where $a_{ij}$ can be either a binary or a non negative real value (weighted matrix). The vector $\vec x =(x_1,....,x_n)'$, containing the eigenvector centrality $x_i$ of node $i$, satisfies
\begin{equation}\label{evec}
A\vec x = \lambda^{max} \vec x
\end{equation}   
i.e. it is a right eigenvector of the adjacency matrix corresponding to the maximal eigenvalue $\lambda^{max}$. Interpreting the network as a representation of a Markov chain with $n$ states and transition probabilities proportional to the weights of the links connecting two nodes (states), it can be seen that the vector $\vec x$ is the stationary probability distribution of the chain. Interestingly the eigenvector centrality is at the heart of Google's PageRank algorithm. In the context of systemic risk, it has been recently modified to define DebtRank \cite{debtrank} for measuring the centrality of the network of mutual exposures and BankRank \cite{bankrank} for measuring the centrality of the bipartite network of banks and assets. In the present paper we will use the original definition of eigenvector centrality of Eq. \ref{evec}.

\subsection{Centrality measures in the interbank multiplex}\label{centralityres}

The aim of this subsection is to compare the centrality measures in different layers of the interbank network. The comparison of the degree of a bank in different layers has been investigated in depth in \cite{Bargigli2014}, also with respect to null models that preserve the degree in average (configuration model). For this reason in the paper we will consider mostly betweenness and eigenvector centrality. 

\begin{figure}[t]
\begin{center}
\includegraphics[width = 0.48\textwidth,keepaspectratio=true]{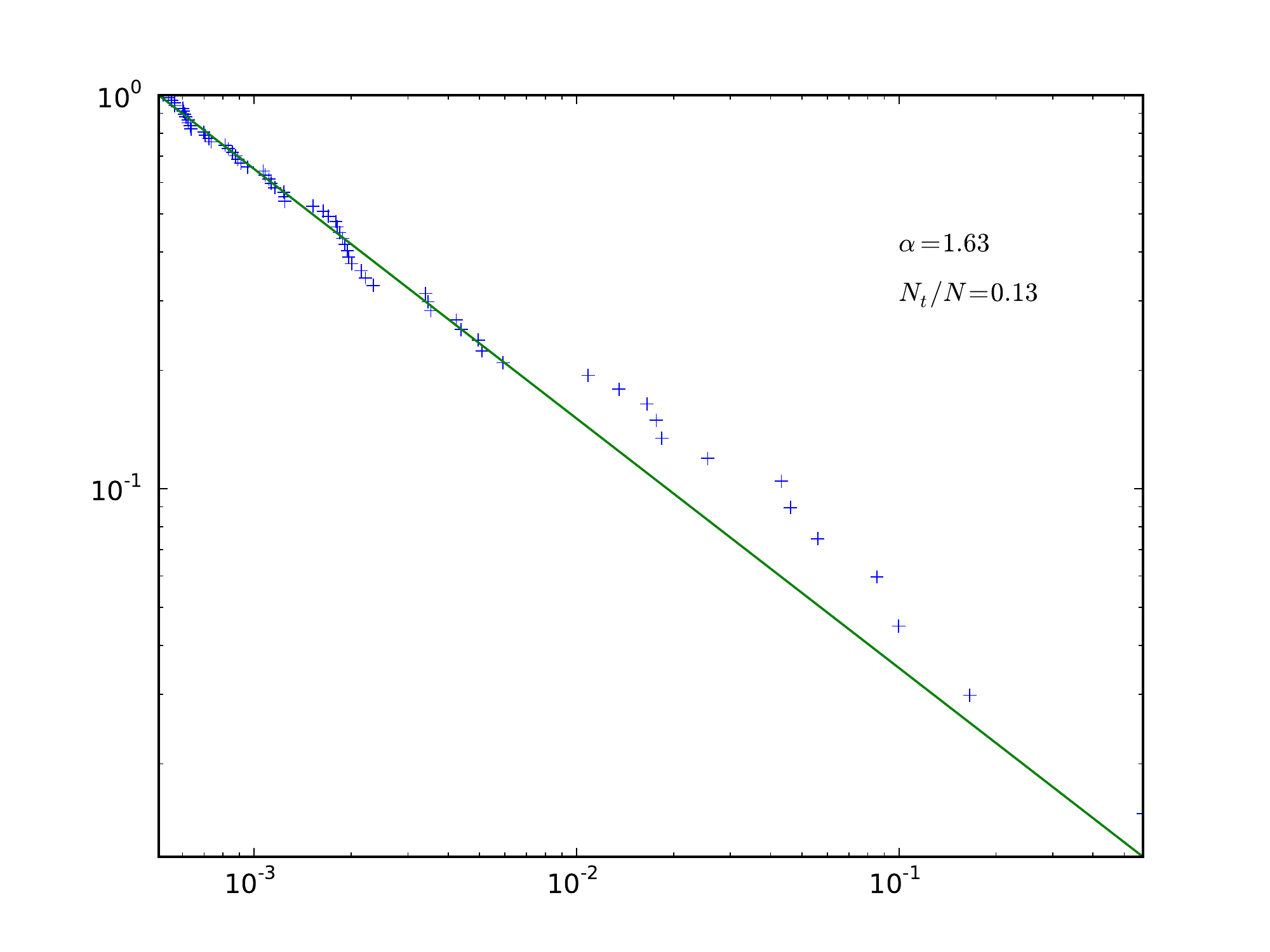}
\includegraphics[width = 0.48\textwidth,keepaspectratio=true]{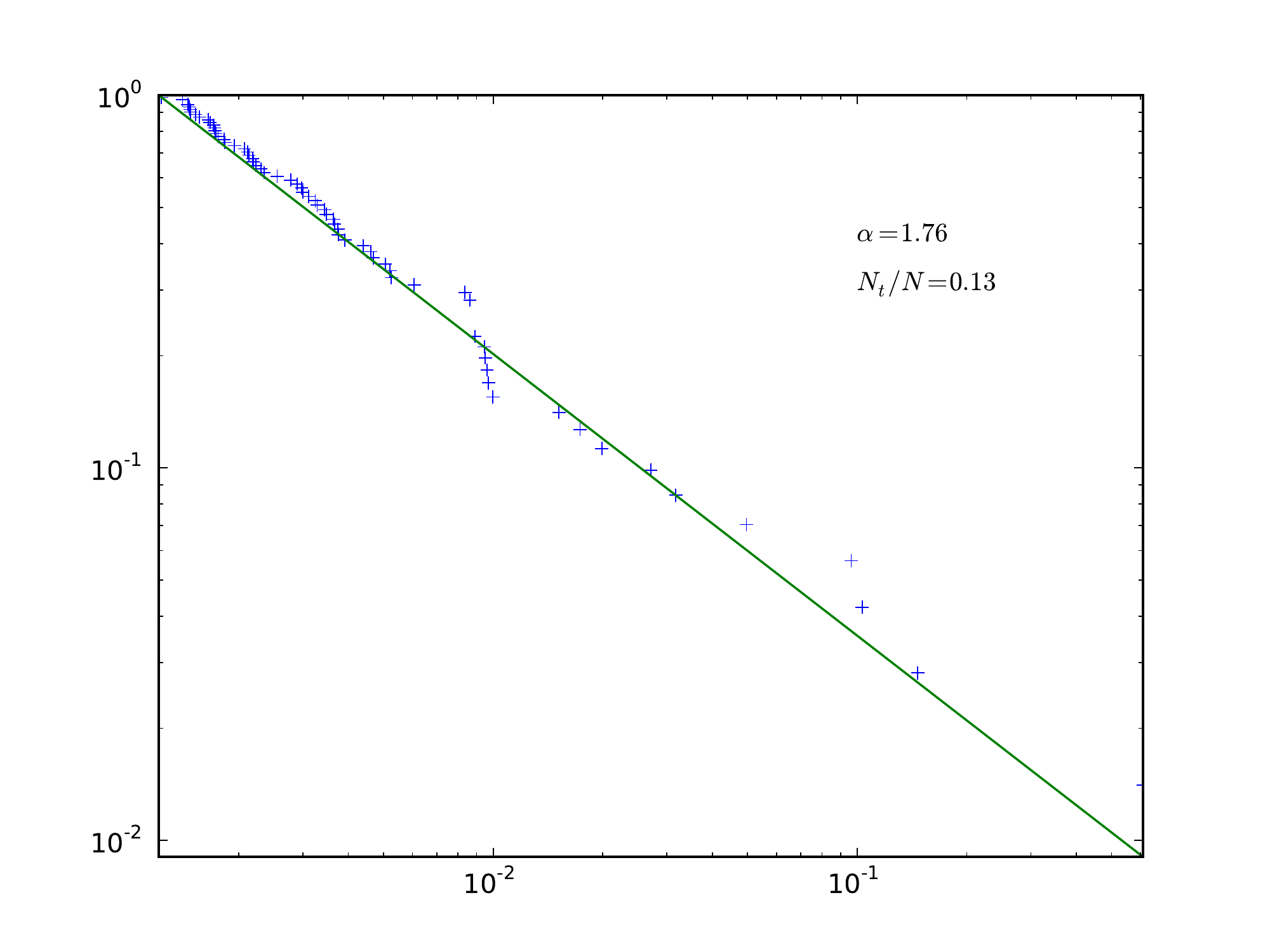}
\end{center}
\caption{Complementary cumulative distribution function of betweenness in the unsecured overnight layer (left panel) and in the  unsecured long term layer (right panel) in 2012. The plot is in double logarithmic scale and the green line is the best fit with a power law function. The tail exponent $\alpha$ is reported in the panels.}
\label{eigdist}
\end{figure}

As a preliminary analysis, figure \ref{eigdist} shows the distribution of betweenness in two important layers, overnight and unsecured medium term in 2012 (similar results are obtained for the other investigated years). The distribution is well fit by a power law function with a tail exponent between $1.5$ and $2$.  This fat tail behavior shows that there is a large heterogeneity of the betweenness centrality 
among the banks in all layers. In part this heterogeneity in the value of the betweenness is due to the scale free behavior of the layers. In fact, as shown in \cite{Bargigli2014}, the degree distribution of each layer is well fit by a power law tail and the estimated values of the tail exponent are remarkably stable across layers and over time, ranging in  $[1.8, 3.5]$ and mostly concentrated around $2.3$. 

\begin{figure}[t]
\begin{center}
\subfloat[out-degree]{\includegraphics[width = 0.4\textwidth,keepaspectratio=true]{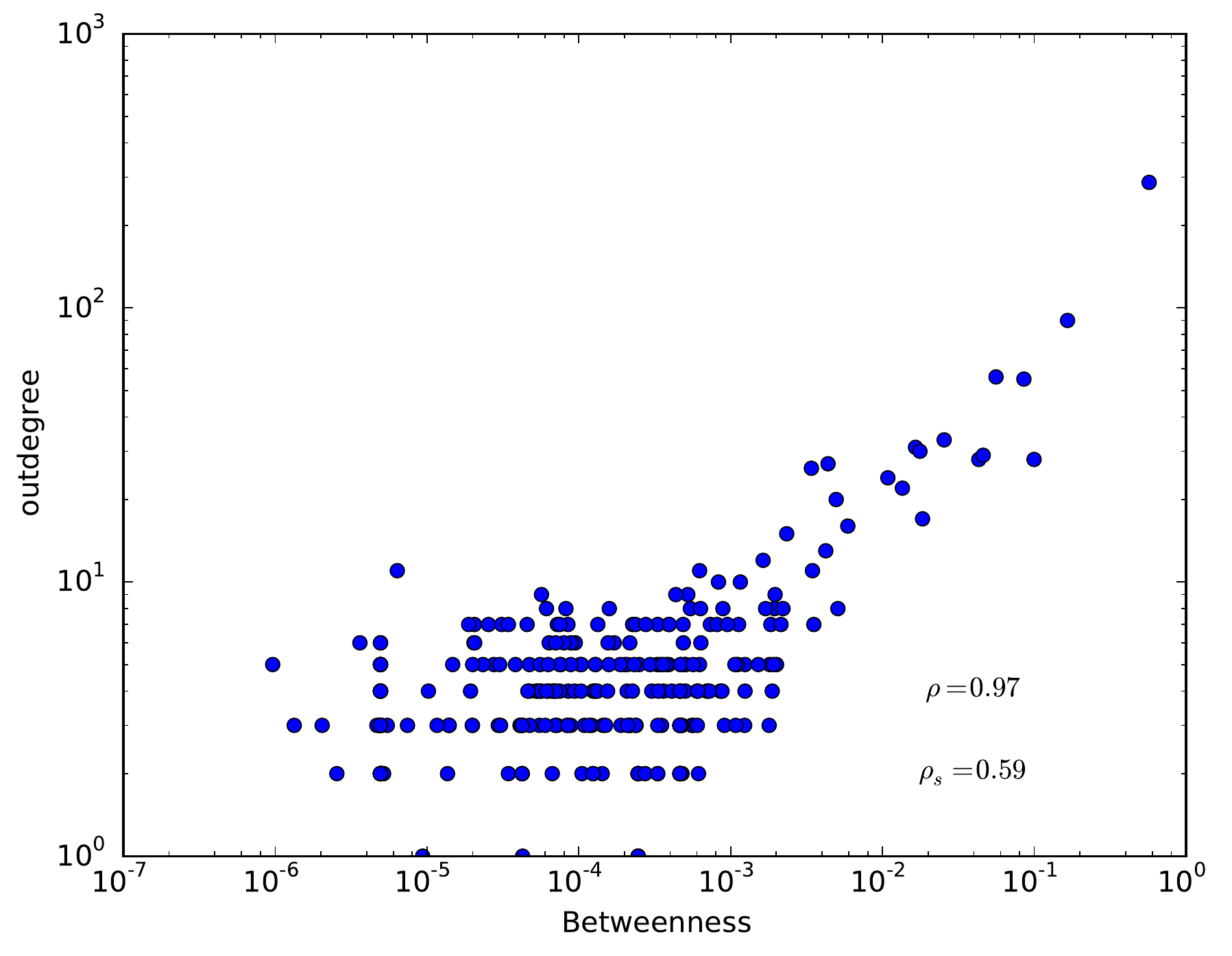}}
\subfloat[in-degree]{\includegraphics[width = 0.4\textwidth,keepaspectratio=true]{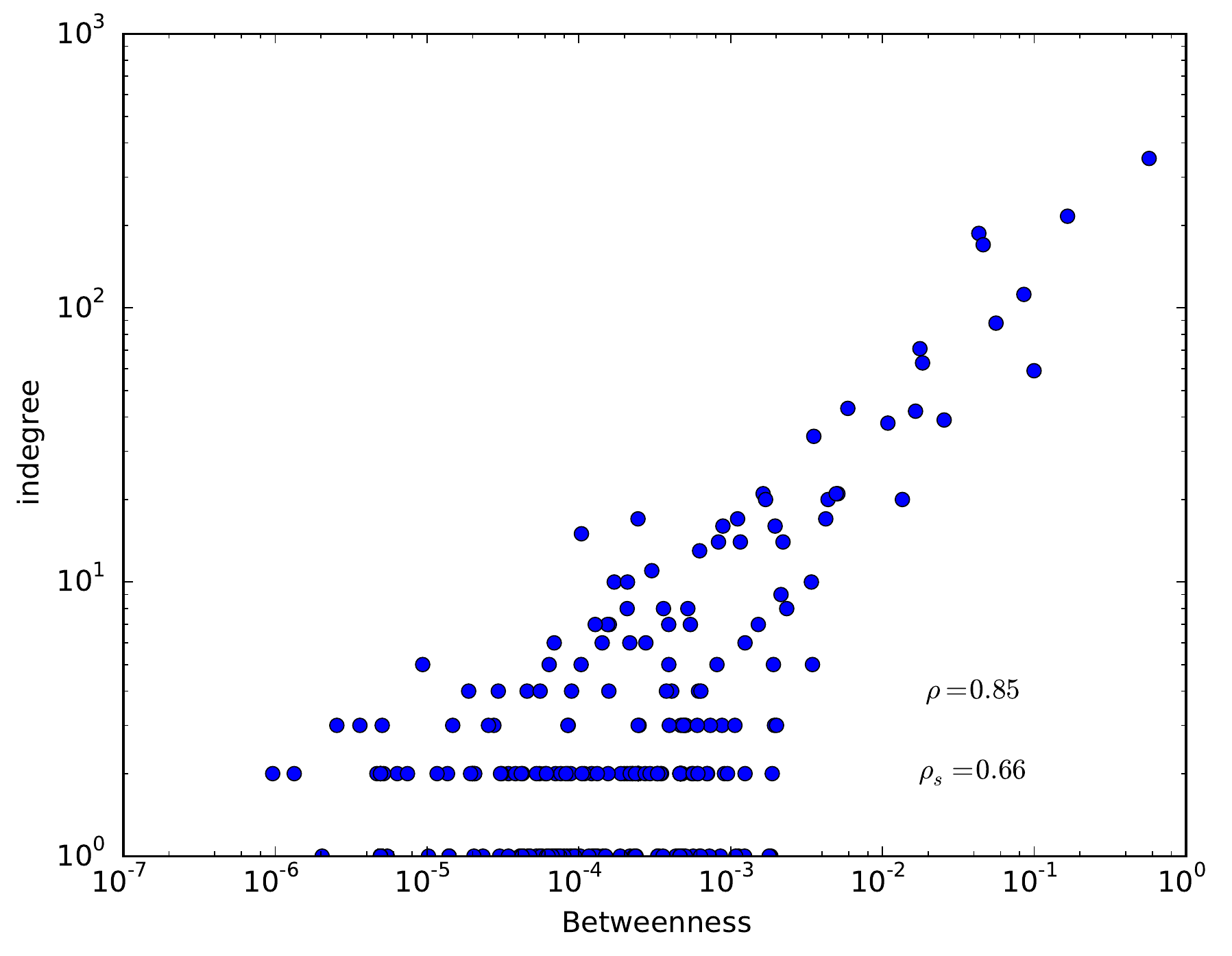}} \\
\subfloat[out-strength]{\includegraphics[width = 0.4\textwidth,keepaspectratio=true]{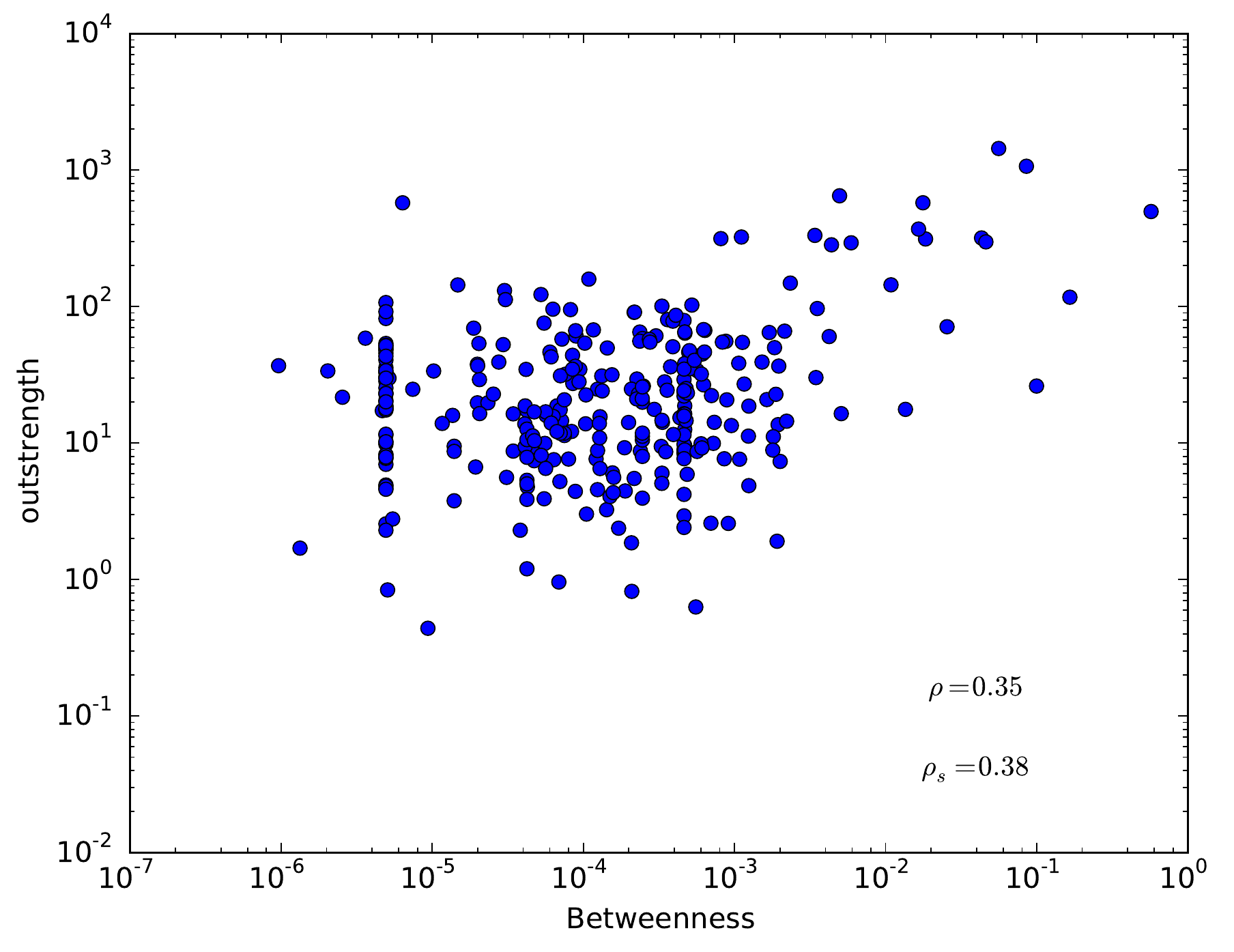}}
\subfloat[in-strength]{\includegraphics[width = 0.4\textwidth,keepaspectratio=true]{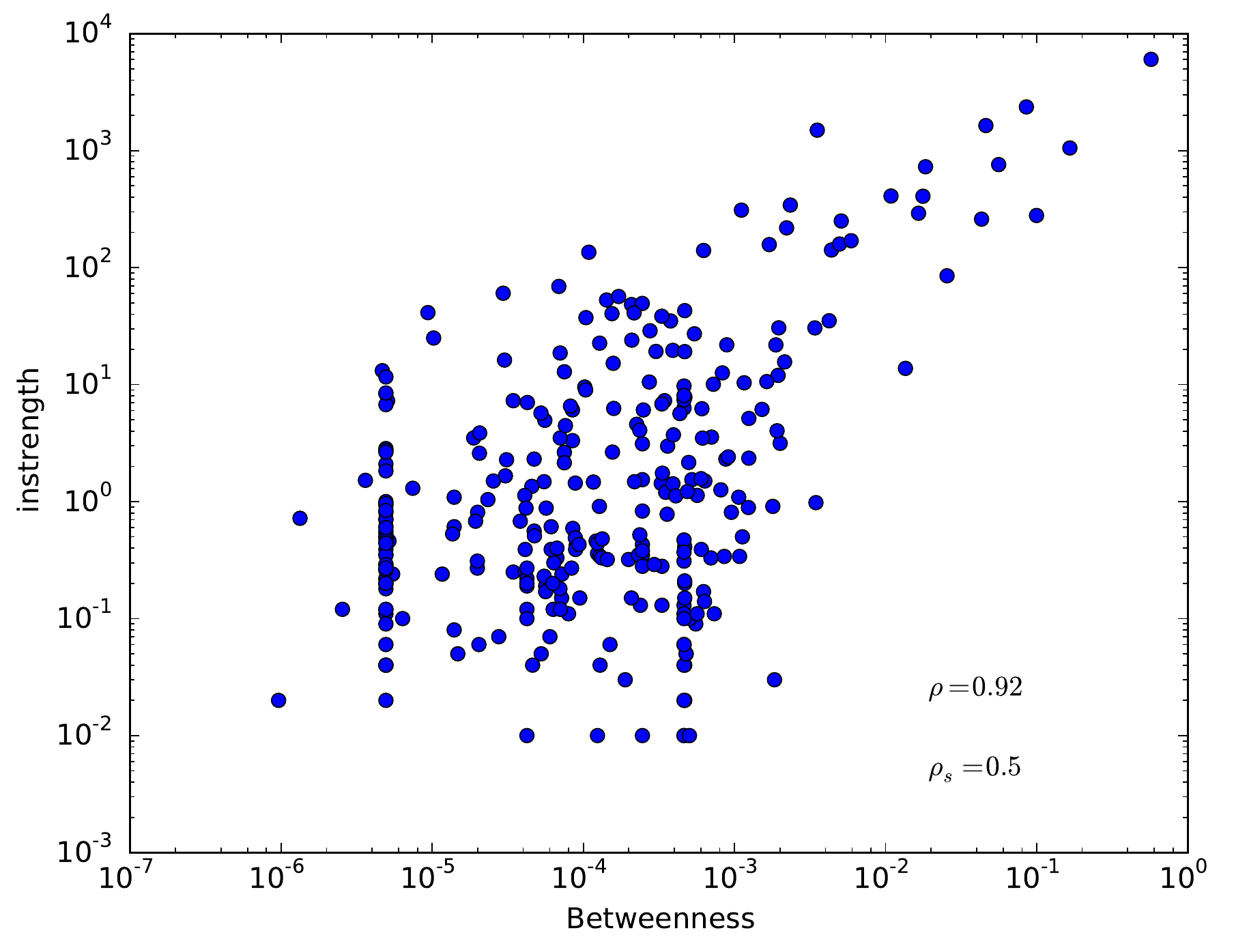}}
\end{center}
\caption{Scatter plot of the betweenness and different nodes properties, namely out-degree (a), in-degree (b), out-strength (c), and in-strength (d). Each point is a bank group. The values of $\rho$ and $\rho_s$ are the Pearson and the Spearman correlation, respectively. Data refer to the unsecured overnight layer in 2012.}
\label{betweennessdegree}
\end{figure}

The relation between degree and betweenness of the banks in the overnight of the Italian Interbank Networks is shown in Fig. \ref{betweennessdegree} (top panels), while the bottom panels show the relation between strength and betweenness. To have a numerical value to quantify the dependence between the two variables, in each figure we report both the value of the Pearson correlation and the value of the Spearman rank correlation. It is worth noticing that in general the former is significantly larger than the latter. This is due to the strong correlation in the right tail of the two variables (i.e. for large and very connected banks). The top panels show that banks with large in- (or out-) degree are also the nodes with high betweenness centrality. The same holds when one considers the in- or out-strength (bottom panels), even if in this case this strong relation is evident only for a smaller interval of large values of strength.  

This analysis shows that large (and/or more interconnected) banks are also typically those more central (as measured by the betweenness)\footnote{As mentioned in Section 3, \cite{Bargigli2014} show that nodes' properties (degree and strength) turn out to be correlated with balance sheet data of banks, in particular with the total assets.
}. Thus, at least in the Italian Interbank Market, it is hard to discriminate between too big to fail and too interconnected to fail. Considering medium and small sized banks, the relation between centrality and degree and especially size becomes significantly noisier, and by looking at figure \ref{betweennessdegree}, it is possible to identify banks with moderate strength but significant centrality. These institutions are likely playing a central role in intermediation in the considered layer or connecting subsets of banks which are only weakly connected.

We then investigate whether banks which are central in a given layer are also central in other layers.  This is important because it gives insight on the degree of specialization of some banks as intermediary for some type of credit. To answer this question we compute centrality measure in different layers and compare the values (or the ranking)  We remind that the number of banks active in the different layers is different \cite{Bargigli2014}. For example, in 2012, of the 533 banks active in the interbank market, 532 had credit relations in the overnight market, 521 were active in the unsecured short term, 450 in the unsecured long term, and 45 in the secured short term. In order to compare layers with different number of nodes, we computed the centrality measures in the whole layer (i.e. including all the banks active in the layer), but we compared the centrality measures only within the subset of banks which were active in both layers. Therefore, when considering rankings, it should be kept in mind that 
these are not absolute rankings, but only rankings among the banks considered in the intersection. 

\begin{figure}[t]
\begin{center}
\subfloat[]{\includegraphics[width = 0.45\textwidth,keepaspectratio=true]{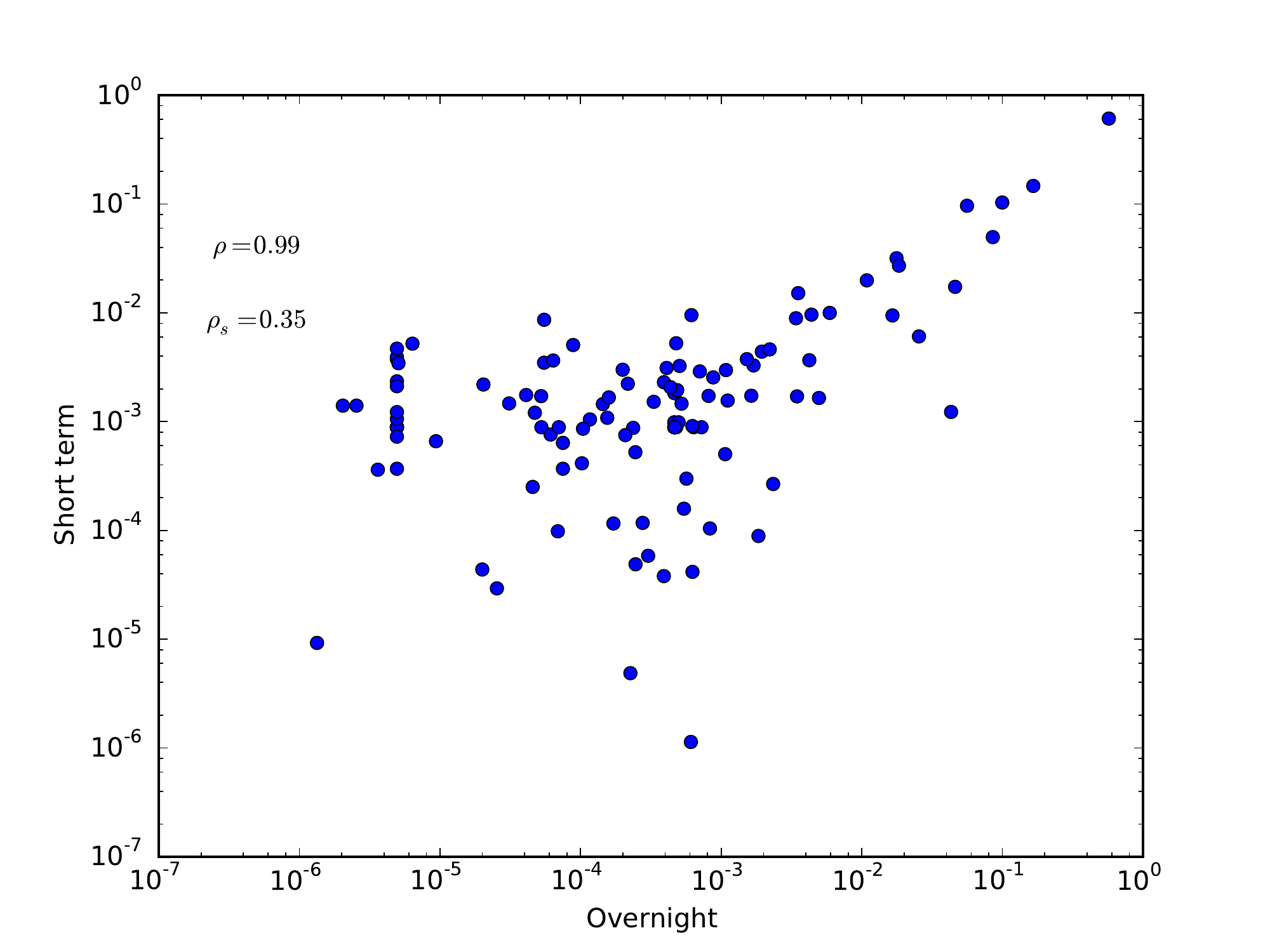}}
\subfloat[]{\includegraphics[width = 0.45\textwidth,keepaspectratio=true]{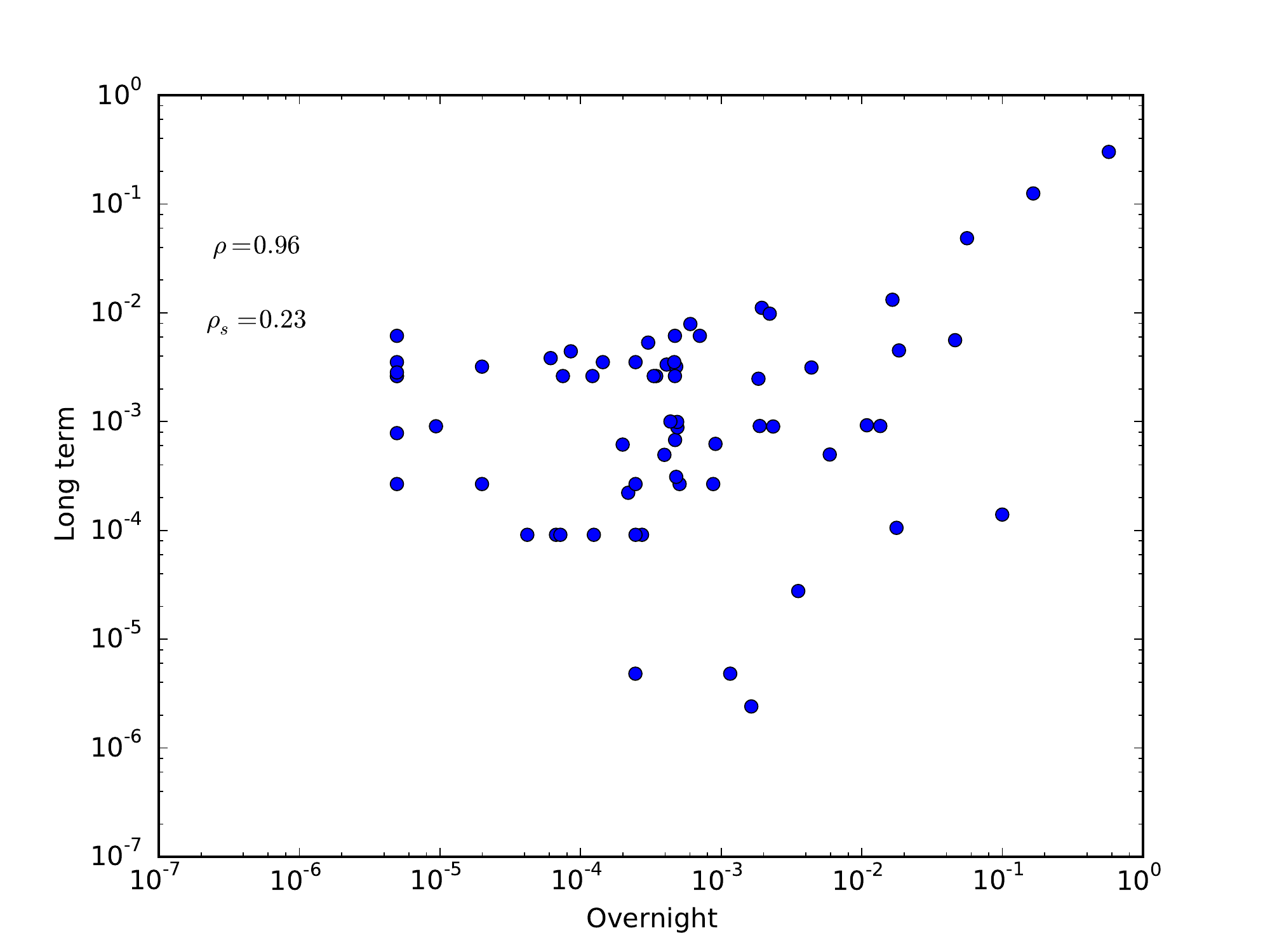}}
\end{center}
\caption{Scatter plot of the betweenness in the unsecured overnight versus that in the unsecured short term layer. The year investigated is 2012. The values of $\rho$ and $\rho_s$ are the Pearson and the Spearman correlation, respectively.}
\label{comparisonbetw}
\end{figure}

Figure \ref{comparisonbetw} shows the scatter plot of the betweenness in the overnight market either versus the betweenness in the unsecured long term (left panel) or versus the betweenness in the unsecured long term (right panel) market. In the figure we also report the value of the Pearson and Spearman correlation and, as explained above, the former is typically much higher of the latter due to the high correlation on the top left part of the scatter plot.

Comparing the short term and overnight layers (left panel), it is evident that top central nodes in one layer are also typically central in the other layer. For medium and low centrality, the correlation is much weaker. On the contrary, the right panel shows that, with the exception of three banks on the top right corner of the figure, there is a very weak correlation between the centrality in the overnight network and in the long term network. This is an indication that in some cases, centrality of a bank, also with respect to the other banks, can be markedly different in different layers. Therefore central banks in the long term layer are not necessarily central in the overnight or in the short term layer.

\begin{figure}[t]
\begin{center}
\subfloat[out-degree]{\includegraphics[width = 0.45\textwidth,keepaspectratio=true]{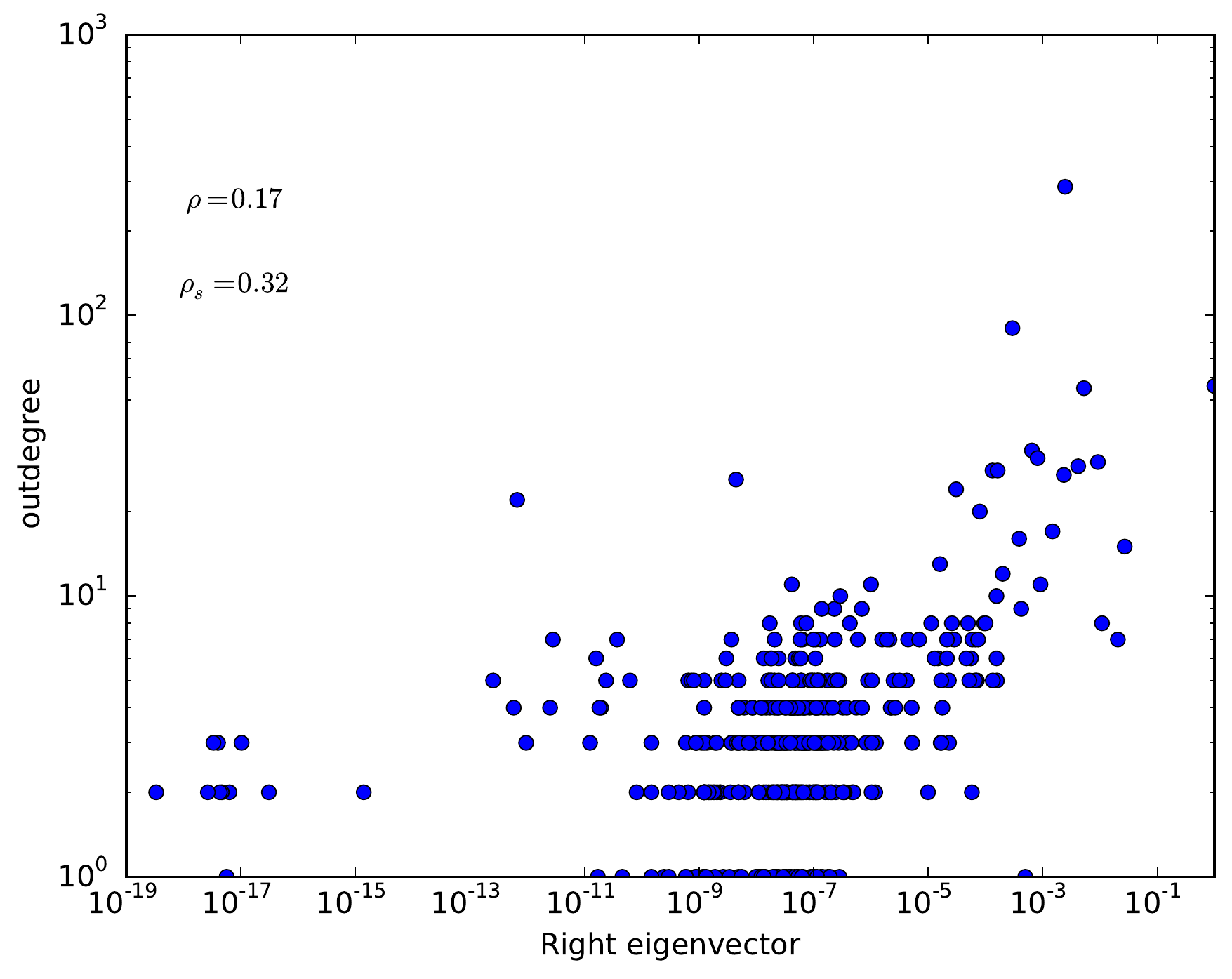}}
\subfloat[in-degree]{\includegraphics[width = 0.45\textwidth,keepaspectratio=true]{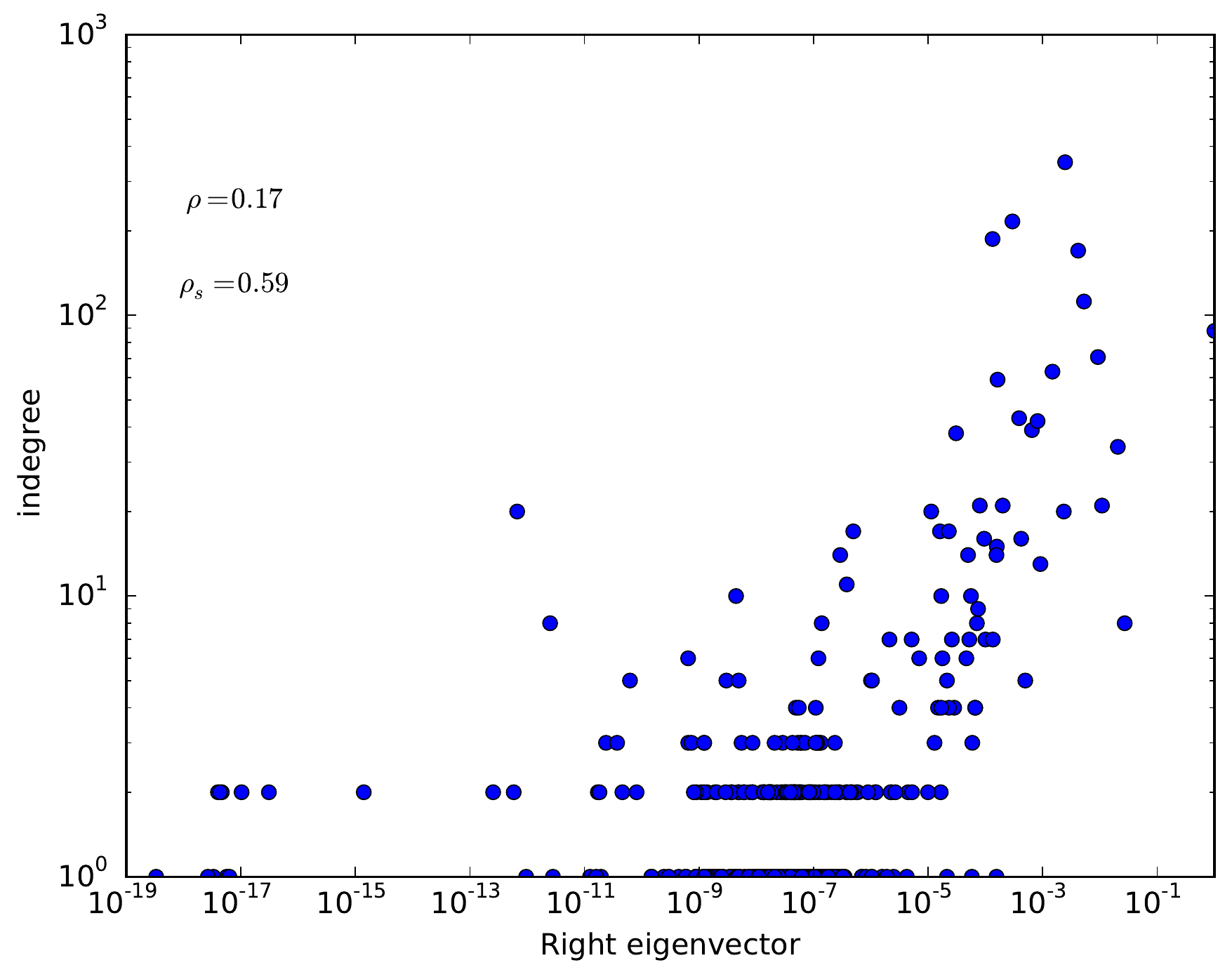}} \\
\subfloat[out-strength]{\includegraphics[width = 0.45\textwidth,keepaspectratio=true]{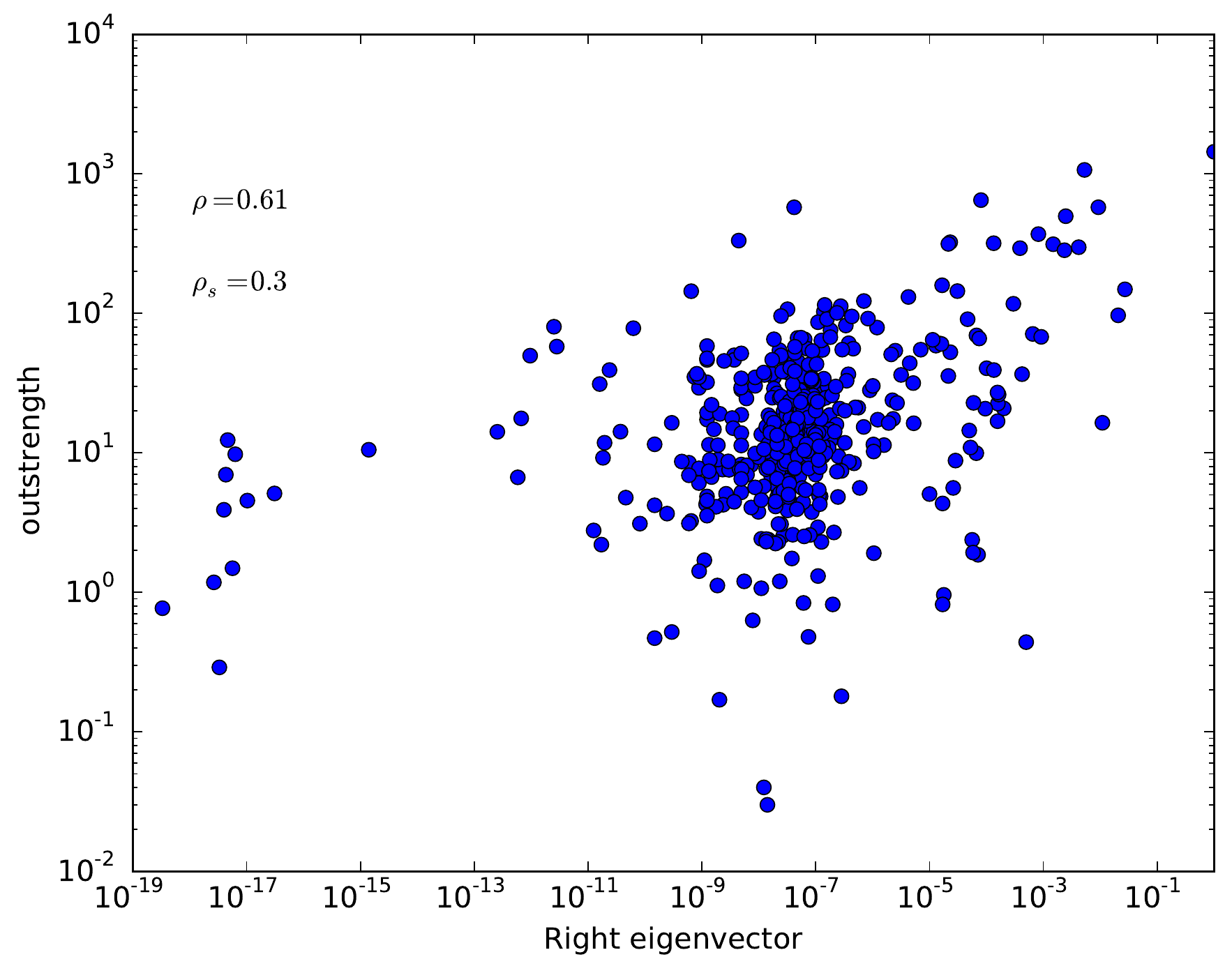}}
\subfloat[in-strength]{\includegraphics[width = 0.45\textwidth,keepaspectratio=true]{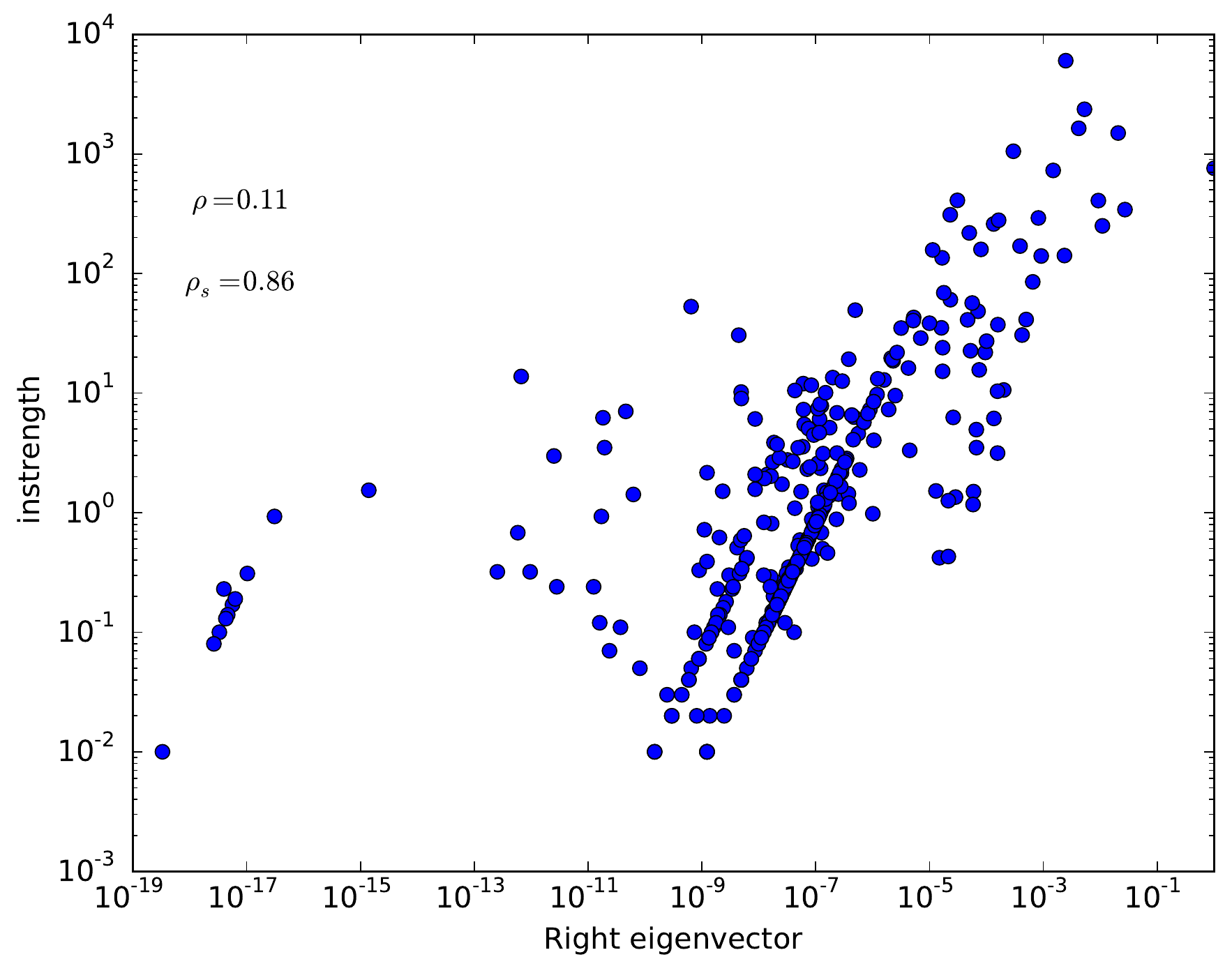}}
\end{center}
\caption{Scatter plot of the eigenvector centrality and different nodes properties, namely out-degree (a), in-degree (b), out-strength (c), and in-strength (d). Each point is a bank group. The values of $\rho$ and $\rho_s$ are the Pearson and the Spearman correlation, respectively. Data refer to the unsecured overnight layer in 2012. }\label{eigscatter}
\end{figure}

Similar conclusions can be drawn by considering the eigenvector centrality. Figure \ref{eigscatter} shows the scatter plot between the eigenvector centrality and in- and out-degree (top panels) and in- and out-strength (bottom panels). The correlation are significantly smaller when compared to those of the betweenness centrality (see Fig. \ref{betweennessdegree}). In this sense eigenvector centrality (and probably DebtRank) brings information on the importance of a bank which is not contained already in the basic measures of degree and strength.

\begin{figure}[t]
\begin{center}
\subfloat[2009]{\includegraphics[width = 0.45\textwidth,keepaspectratio=true]{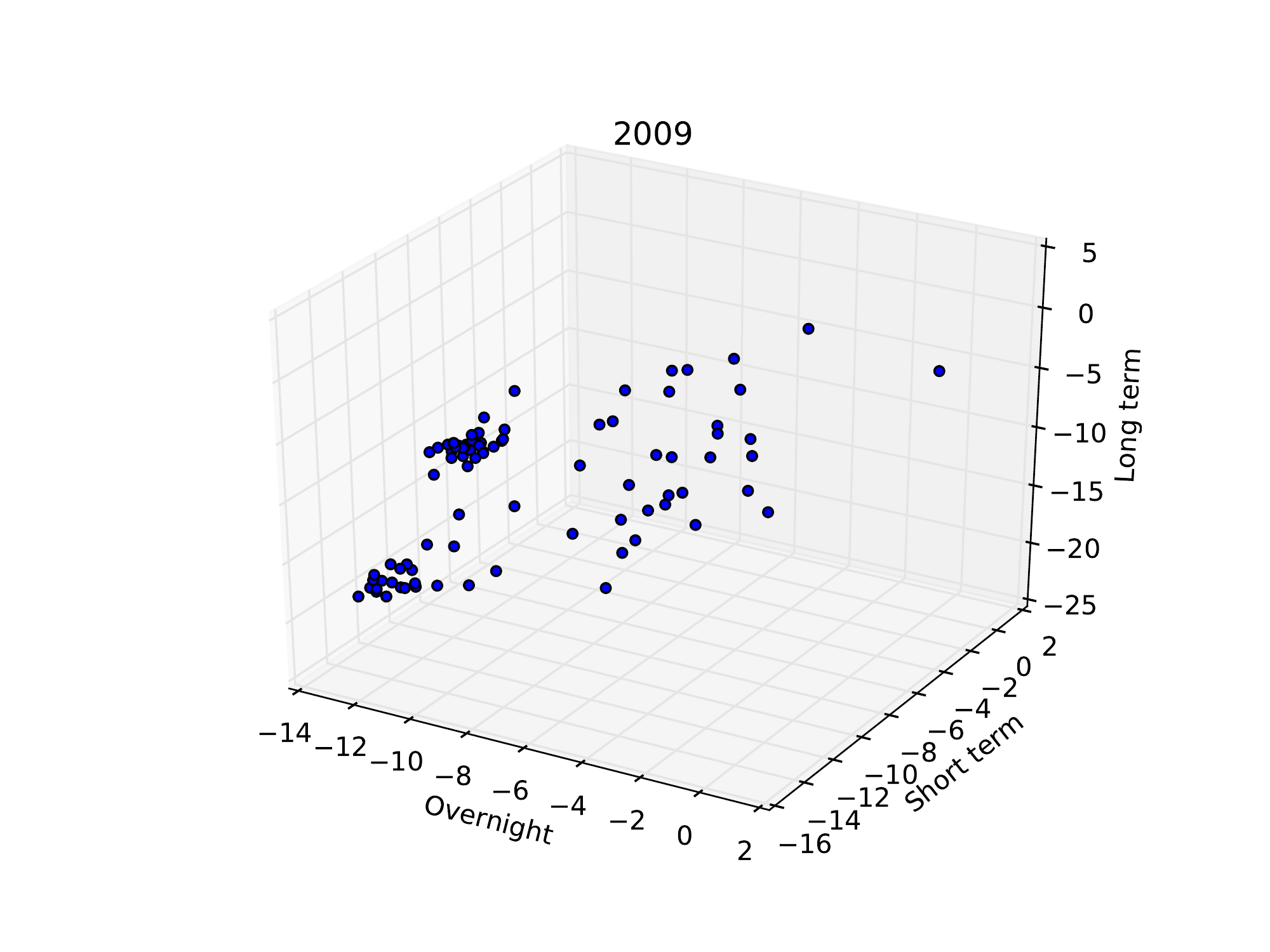}}
\subfloat[2010]{\includegraphics[width = 0.45\textwidth,keepaspectratio=true]{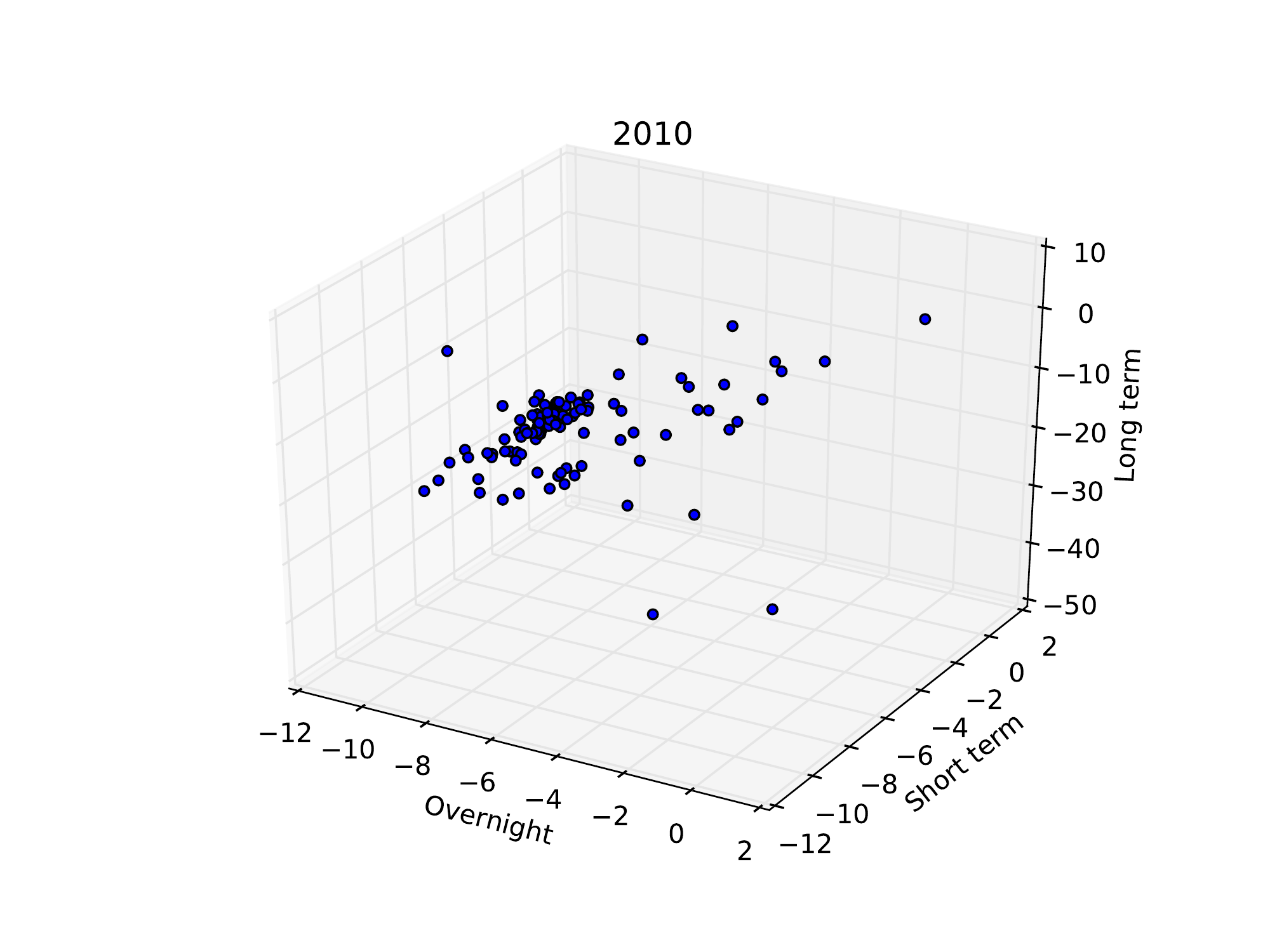}}\\
\subfloat[2011]{\includegraphics[width = 0.45\textwidth,keepaspectratio=true]{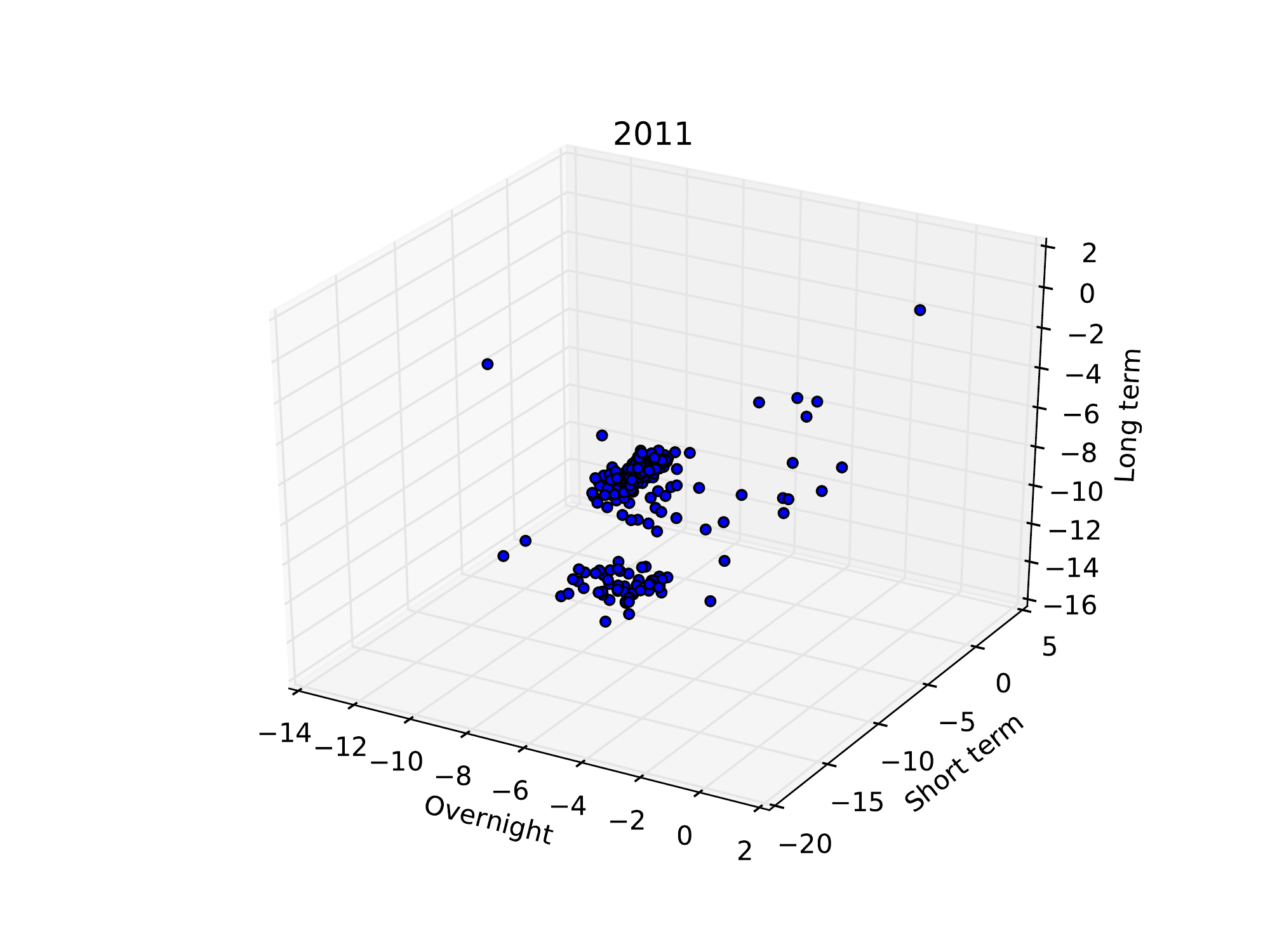}}
\subfloat[2012]{\includegraphics[width = 0.45\textwidth,keepaspectratio=true]{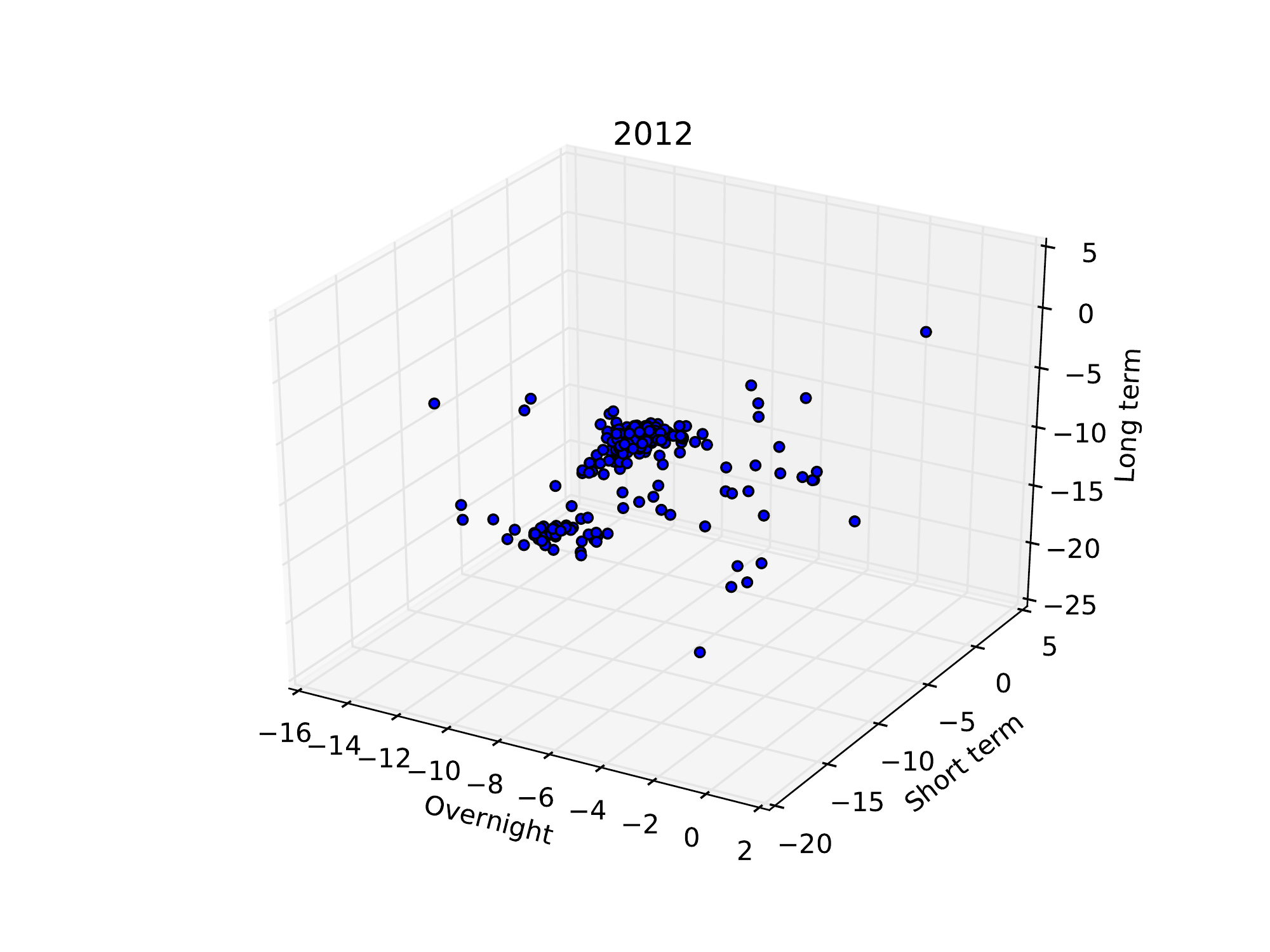}}
\end{center}
\caption{Three dimensional scatter plot of the decimal logarithm of the eigenvector centrality in the three unsecured layers in the period 2009-2012.}\label{3d}
\end{figure}

Finally, in figure \ref{3d} we show the three dimensional plots of the eigenvector centrality in the three unsecured layers, namely overnight, short term, and long term. We show the results for the four years because we find a different behavior in the different years.  In all years we find a very large dispersion of points, indicating that the eigenvector centrality in the three considered layers is very different. Thus the importance of a node, as measured by the eigenvector centrality, is typically quite layer specific.  

\subsection{Comparison with null models}\label{null}

As a last analysis, we compare our findings on the centrality metrics with suitable statistical null models. In the last two decades complex network theory has introduced a large number of metrics able to capture many interesting aspects of the organization of networks, such as clustering, assortativity, core-periphery structure, etc. However, from a statistical point of view, it is not always clear which of these properties carry some information which is not already contained, in some form, in simpler properties of the network. For example, one could ask whether the core-periphery organization observed in many networks (included the interbank ones) is not a mere consequence of the large heterogeneity in degree of the considered network. In fact, in a network with heterogeneous degree distribution, a core periphery emerges even if the links among nodes are assigned randomly.

To properly  answer to these questions one needs to build statistical models of networks, which allow computing a probability distribution of graphs. Such a distribution is calibrated on the investigated real network, and it is chosen in such a way to preserve some low order properties (e.g. the degree of each node). Then one computes, analytically or computationally, the distribution of the considered high order property (e.g. the size of the core) and extracts from it  a $p$-value for the value observed in the real data. 

One of the most common methods for building null models is by using the Maximum Entropy Principle.\footnote{See for instance \cite{bar2013, bar2011396, 2011arXiv1112.2895F, RePEc:dnb:dnbwpp:348, mistrulli2011assessing, park2004statistical, 1367-2630-13-8-083001}.} One looks for the probability distribution for graphs, $P(G)$, which maximizes the Shannon entropy 
\begin{equation}
S[P(G)]=-\sum_G P(G) \ln P(G)
\end{equation}
under suitable constraints, including the normalization $\sum_G P(G)=1$.  More details on the construction and estimation of Maximum Entropy models for networks are reviewed in Appendix 4 of \cite{Bargigli2014}.

Here we consider the so called Directed Binary Configuration Model (DBCM) (see \cite{Bargigli2014}). In this case we impose $2n$ constraints, namely the average in- and out-degree of each node. This model can be fitted from a real network using Maximum Likelihood and one can then sample from a graph distribution of a DBCM calibrated on the a real network.  We can compute the distribution of the betweenness centrality for each node of the network under the DBCM. We then compute the $p$-value of the betweenness centrality of each node observed in the real network. 

\begin{figure}[t]
\begin{center}
\subfloat[Overnight]{\includegraphics[width = 0.45\textwidth,keepaspectratio=true]{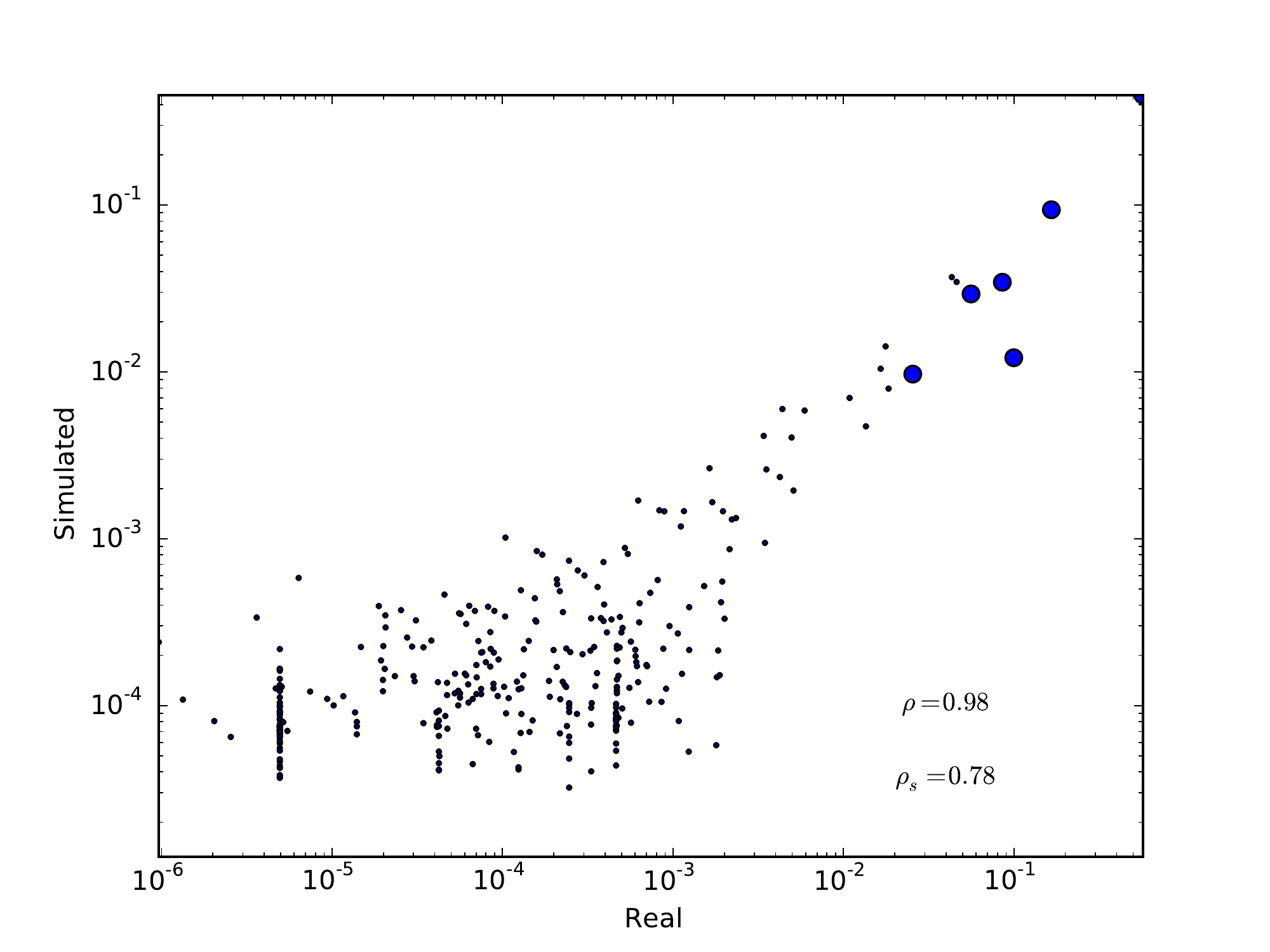}}
\subfloat[Short term]{\includegraphics[width = 0.45\textwidth,keepaspectratio=true]{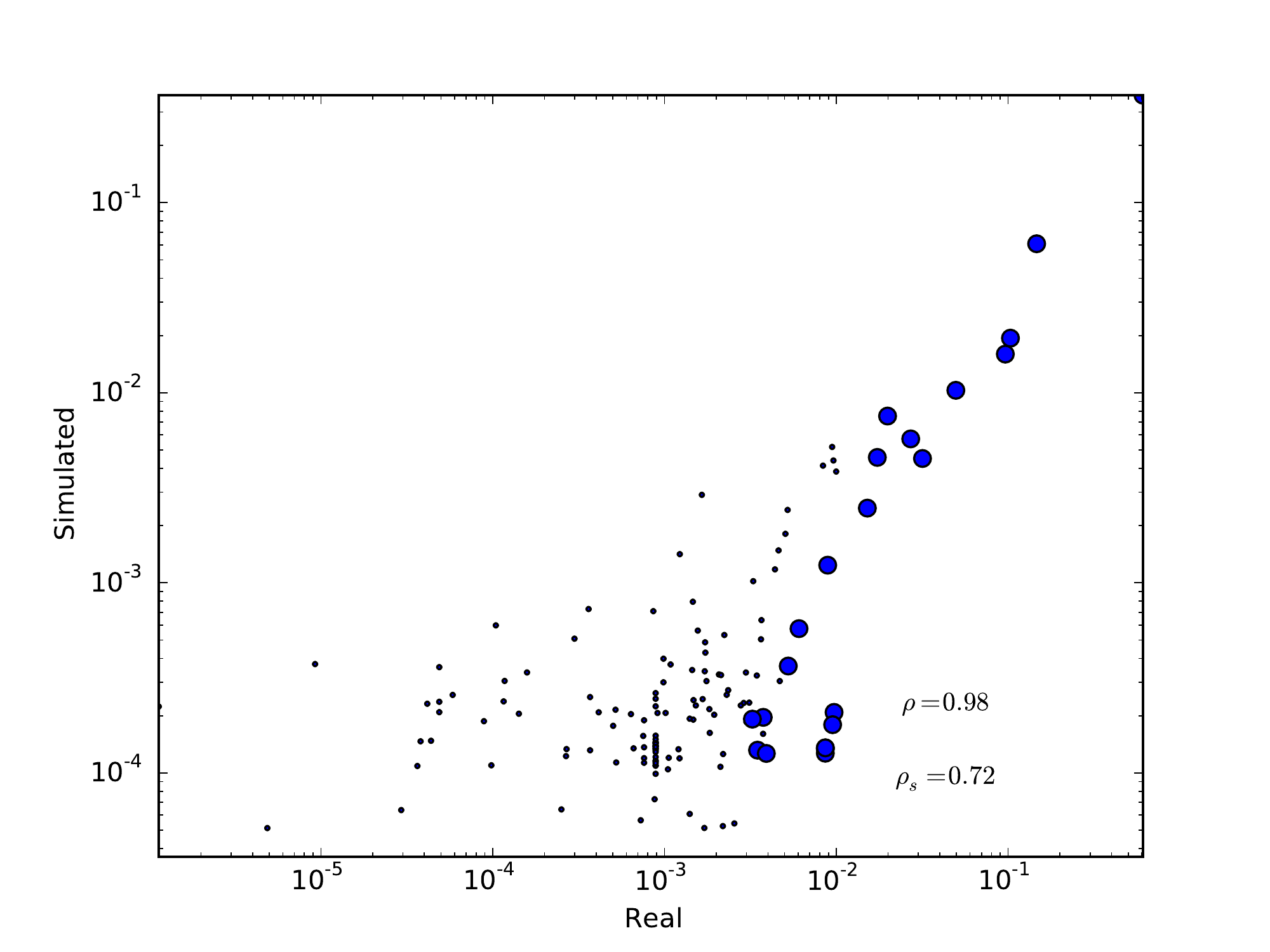}}\\
\subfloat[Long term]{\includegraphics[width = 0.45\textwidth,keepaspectratio=true]{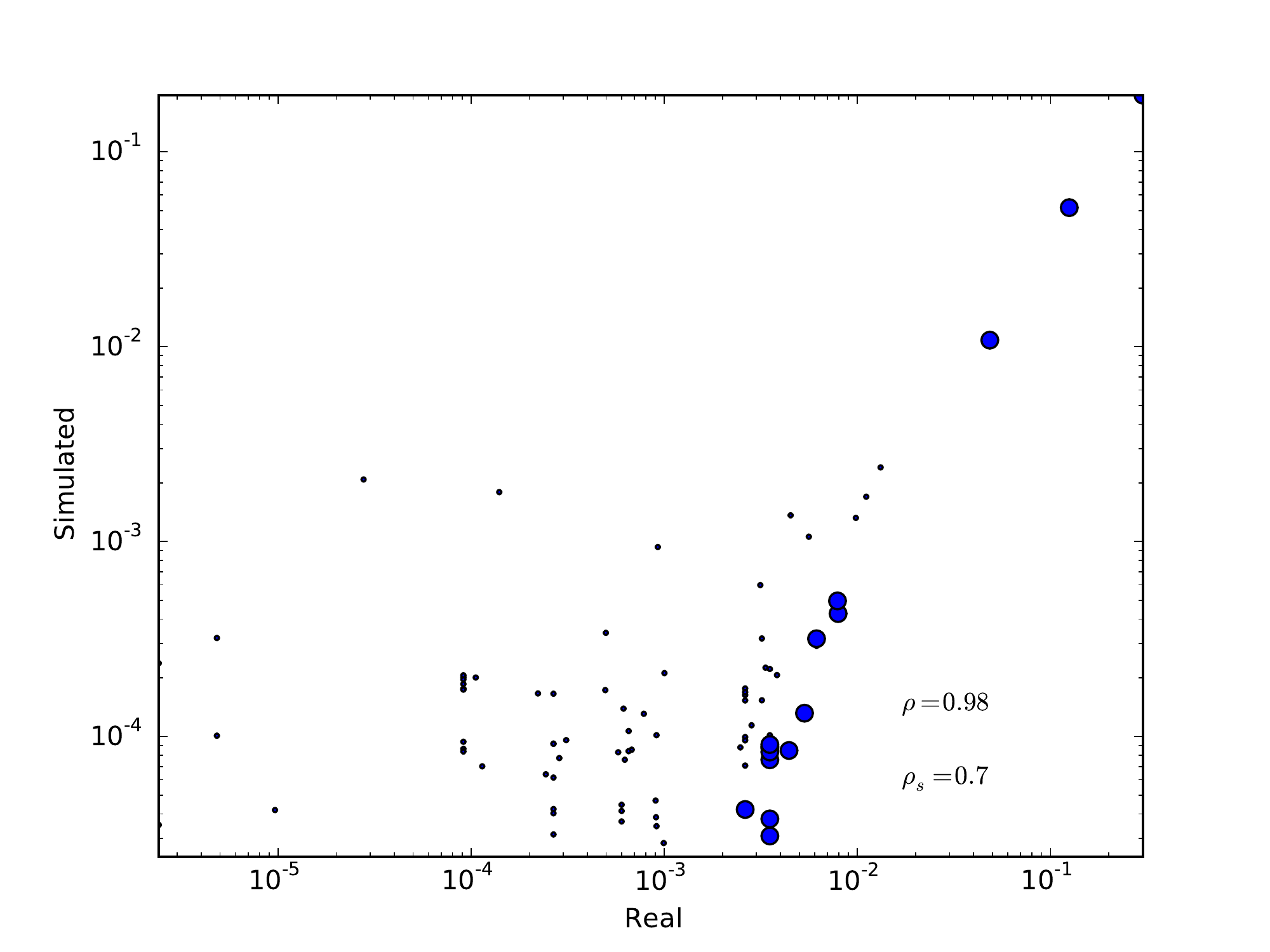}}
\end{center}
\caption{Scatter plot of the betweenness centrality in the real network and in the one simulated from the Maximum Entropy (ME) ensemble corresponding to the directed configuration model.  Larger circles correspond to nodes for which the (ME) null model is rejected with $1\%$ confidence.  The values of $\rho$ and $\rho_s$ are the Pearson and the Spearman correlation, respectively. Data refer to 2012.} \label{nullmodels}
\end{figure}

Figure \ref{nullmodels} shows the  scatter plot of the betweenness centrality in the real network versus the one simulated in the DBCM. As a first observation we note that there is a significantly high Pearson and Spearman correlation between the betweenness in the real and simulated network. In this sense we can conclude that a significant fraction of the betweenness centrality property of a node is a consequence of the degree distribution. This observation is in agreement with the empirical observation described above that betweenness centrality is very correlated with degree. However deviations from the null model can be identified by computing for each node the $p-$value, comparing the centrality in the real network with the values obtained by a large sample of network realization. In figure \ref{nullmodels} larger circles correspond to nodes for which the (ME) null model is rejected with $1\%$ confidence. Interestingly, the nodes for which betweenness centrality is larger than what expected from their 
degree (and the Maximum Entropy null model) are mostly large banks. 

\section{Conclusions}\label{conclusions}

This paper reviews the multiplex description of the interbank market proposed in \cite{Bargigli2014}. In this description, each layer is a network of interbank exposures characterized by the maturity and the presence of a collateral. By using an unique dataset of the Italian Interbank market, we discuss the main findings related to (i) the similarity of the topological properties of the different layers, (ii) the link-by-link similarity of pairs of layers and (iii) the use of Maximum Entropy approach for the construction of layer-specific null models.

The main original contribution of the paper is the investigation of two centrality measures, namely betweenness and eigenvector centrality, in the multiplex describing the Italian Interbank market. We found that the correlation between the centrality of a bank in different layer is significant, but not extremely large. There are several medium sized banks which are central in some layers and peripheral in others. Very large banks are typically central in all layers, but this can be due - at least partially - to the constraint given by their typically large degree. Interestingly, the centrality of large banks is not explained by Maximum Entropy null models preserving the degree of each node. This indicates deviations from centrality measure expected by the degree distribution. Finally, betweenness centrality is more correlated with degree than eigenvector centrality. This finding seems to indicate that the eigenvector centrality is less affected by information already contained in the degree, strength, and 
bank size.

\begin{acknowledgement}
The views expressed in the article are those of the authors only and do not involve the responsibility of the Bank of Italy. This paper should not be reported as representing the views of the European Central Bank (ECB). The views expressed are those of the authors and do not necessarily reflect those of the ECB. The research leading to these results has received funding from the European Union, Seventh Framework Programme FP7/2007- 2013 under grant agreement CRISIS-ICT-2011-288501 and from the INET-funded grant ‘New tools in Credit Network Modeling with Heterogeneous Agents’. L. Infante contributed while visiting the NYU Department of Economics.\end{acknowledgement}

\end{document}